\documentclass[useAMS,aasms,astrobib,usenatbib]{mn2e}
\usepackage{epsfig,amssymb,graphicx,rotating,color}
\usepackage{hyperref}
\usepackage{txfonts}
\title{The curiously circular orbit of Kepler-16b}
\author[Dunhill \& Alexander]{A.C. Dunhill\thanks{E-mail: alex.dunhill@leicester.ac.uk} and R.D. Alexander
\\Department of Physics \& Astronomy, University of Leicester, Leicester, LE1 7RH}
\begin{document}
\voffset=-0.5in

\newcommand{\Msun}{$M_{\odot}$}
\newcommand{\Mjup}{$M_{\mathrm{Jup}}$}
\newcommand{\mockalph}[1]{}
\newcommand{\sigmin}{$10$ g cm$^{-2}$}

\date{Accepted 2013 August 1.  Received 2013 July 25; in original form 2013 May 28}

\pagerange{\pageref{firstpage}--\pageref{lastpage}} \pubyear{2013}

\maketitle

\label{firstpage}

\begin{abstract}
The recent discovery of a number of circumbinary planets lends a new tool to astrophysicists seeking to understand how and where planet formation takes place. Of the increasingly numerous circumbinary systems, Kepler-16 is arguably the most dynamically interesting: it consists of a planet on an almost perfectly circular orbit ($e = 0.0069$) around a moderately eccentric binary ($e = 0.16$). We present high-resolution 3D smoothed-particle hydrodynamics simulations of a Kepler-16 analogue embedded in a circumbinary disc, and show that the planet's eccentricity is damped by its interaction with the protoplanetary disc. We use this to place a lower limit on the gas surface density in the real disc through which Kepler-16b migrated of $\Sigma_{\mathrm{min}} \sim$ \sigmin. This suggests that Kepler-16b, and other circumbinary planets, formed and migrated in relatively massive discs. We argue that secular evolution of circumbinary discs requires that these planets likely formed early on in the lifetime of the disc and migrated inwards before the disc lost a significant amount of its original mass.
\end{abstract}


\section{Introduction}

Most Sun-like stars form in multiple systems \citep*[e.g.][]{duquennoymayor91,ghezetal93,leinertetal93,simonetal95}. The question of how planet formation in these systems differs from single stars is therefore a pertinent one, and has been the subject of much recent theoretical and numerical work. This has considered both S-type circumstellar orbits \citep[planets orbiting only one star in the binary; e.g.][]{kleynelson08,mullerkley12} and P-type circumbinary orbits \citep*[e.g.][]{paardekooperetal08,pierensnelson08,marzarietal08,fragneretal11}.  While it seems that planet formation is relatively uninhibited in the former case, at least for relatively wide binary separations \citep[see e.g. $\alpha$ Cen Bb, an earth-mass planet in a circum-secondary orbit;][]{dumusqueetal12}, the latter poses a significant challenge. In particular, planetesimal growth at small semimajor axis (within a factor of a few of the binary semimajor axis) is strongly inhibited by the large velocity dispersion induced by the central binary \citep{paardekooperetal08,marzarietal08}.

The recent detection of a number of planets orbiting main-sequence binary stars by the \textit{Kepler} mission \citep{doyleetal11,welshetal12,oroszetal12a,oroszetal12b,schwambetal13} challenges this theoretical understanding. All of the six systems reported to date contain planets in relatively short-period orbits around close eclipsing binaries (binary semimajor-axes $a_{\mathrm{b}} < 0.25$ au, planet semimajor-axes $a_{\mathrm{p}} \lesssim 1$ au). While transit surveys such as \textit{Kepler} are naturally biased towards finding short-period systems first, it is noteworthy that these planets are predominantly found just on the edge of dynamical stability \citep{holmanwiegert99}.  Moreover, the fraction of close binaries hosting circumbinary planets is estimated to be $\gtrsim 1$--10 per cent \citep{welshetal12}, implying that planets form readily in circumbinary discs.

Four of the seven of the circumbinary planets discovered by {\it Kepler} to date have low eccentricities, $e_{\mathrm{p}} < 0.05$. Of the exceptions, Kepler-34b and PH1 both orbit eccentric binaries ($e_\mathrm{b} \gtrsim 0.2$), while Kepler-47c is the outer planet in a two-planet system and is therefore a special case; the remainder are in near-circular orbits around low- or moderate-eccentricity binaries.  At close separations circumbinary orbits are highly non-Keplerian, so directly measuring their eccentricities does not necessarily give a clear picture of the orbits. The planet's eccentricity $e_{\mathrm{p}}$ comprises the sum of two components, referred to as the forced and free eccentricities ($e_{\mathrm{forced}}$ and $e_{\mathrm{free}}$). The forced eccentricity is driven by the potential of the central binary, and only the free eccentricity is a parameter of the planet's orbit alone \citep{murraydermott99}. However, an analysis of the orbits of Kepler-16b, Kepler-34b and Kepler-35b by \citet{leunglee13} found that $e_{\mathrm{free}}$ is an order of magnitude higher for Kepler-34b than the others, consistent with its higher measured eccentricity.

Of particular interest is Kepler-16, the first of these systems discovered. It is a closely packed system and is aligned to a very high degree: the planes of the binary and planetary orbits, are aligned to within $0.3^{\circ}$ \citep{doyleetal11} and the binary orbit is aligned to the spin of the primary star to within $3^{\circ}$ \citep{winnetal11}. The total stellar mass is approximately 0.9\Msun, with a mass ratio of 0.3, and the binary has eccentricity $e_{\mathrm{b}} = 0.16$.  The planet Kepler-16b is approximately the mass of Saturn \citep[0.333\Mjup][]{doyleetal11,benderetal12}, with a measured eccentricity of $e_{\mathrm{p}} = 0.0069$; \citet{leunglee13} find that $e_{\mathrm{free}} = 0.03$.

Since its discovery, a number of papers have discussed the formation of this planet. \citet{paardekooperetal12} argued that Kepler-16b is unlikely to have formed in situ, due to the high planetesimal velocity dispersion induced by the binary. \citet{meschiari12} found similar results, and while \citet{rafikov12} showed analytically that a massive disc could damp the velocity dispersion, this effect is probably not sufficient to allow Kepler-16b to form in situ. Indeed, numerical simulations by \citet{marzarietal13} suggest that in practice disc self-gravity may have the opposite effect, again inhibiting planetesimal growth.

The broad conclusion from these papers is that Kepler-16b cannot have formed in its current orbit, and must presumably have formed at larger radius and migrated inwards. This picture is consistent with the highly aligned nature of the system, as well as the fact that it is at the very edge of stability -- the planet is located at the natural inner edge for a circumbinary disc. However, this picture of gentle disc-driven migration is complicated by the work of \citet{pierensnelson08}, who found that a Saturn-mass planet migrating through such a disc is likely to attain a significant eccentricity. Although the measured orbital parameters are expected to osculate under the influence of the binary, \citet{leunglee13} have shown that the maximum eccentricity is still low ($e_{\mathrm p}<0.07$). Moroever, \citet{popovashevchenko12} find that the planet's eccentricity cannot ever have been much greater than $e_{\mathrm{p}} = 0.05$ without its orbit becoming unstable.

These arguments point very strongly towards a picture where Kepler-16b's orbit has never been significantly eccentric, which in turn suggests that its free eccentricity was damped to near zero due to the interaction with its parent protoplanetary disc.  In order to investigate this further, we have carried out high resolution smoothed particle hydrodynamics (SPH) simulations of an analogue to the Kepler-16 system embedded in a circumbinary disc, seeking to characterise how the disc would have affected the angular momentum of the planet (and thus its eccentricity) as it migrated. In section \ref{method} we describe the numerical method used, and present the results of our simulations in section \ref{results}. We discuss the implications of these results in section \ref{discussion} and outline our conclusions in section \ref{conclusions}.

\section{Method}\label{method}
We follow the method of \citet*{dunhilletal13}, using a modified version of the hybrid N-body/SPH code \textsc{Gadget-2} \citep{springel05}. The code has been adapted so that it calculates gravity for the N-body particles via direct summation, and uses an explicit Navier-Stokes viscosity on top of a \citet{morrismonaghan97} artificial viscosity. The Navier-Stokes viscosity is paramaterized as a \citet{shakurasunyaev73} $\alpha$-disc, following the method of \citet{lodatoprice10}. This method allows us to explore discs with $\alpha \gtrsim 0.005$ \citep{dunhilletal13}, and in the simulations described below we chose $\alpha = 0.01$.

We adopt a system of code units such that the unit of distance is $0.22431$ au \citep[equal to the binary semimajor-axis;][]{doyleetal11}. The unit of mass is equal to the sum of the masses of the the binary components and the planet, and the time unit is set to $41.08$ days, the orbital period of the binary. This sets the gravitational constant $G = 4\pi^2$ in code units. The planet and the binary components were modelled as N-body particles, and thus their orbital elements were free to evolve throughout the simulations.

\subsection{Disc model}\label{disc model}
Using this code we have performed simulations of a Kepler-16 analogue system embedded in a circumbinary disc. The discs were modelled using $10^7$ SPH particles. Initial particle positions in the disc plane were generated by randomly sampling the distribution 
\begin{equation}
\Sigma(R) = \Sigma_{\mathrm{p}}\,({R}/{a_{\mathrm{p}}})^{-\gamma},
\label{eq_sigma1}
\end{equation}
where $\Sigma$ is the surface density, $\Sigma_{\mathrm{p}} = 100$ g cm$^{-2}$ is the surface density at the radius of the planet's orbit, $R$ is the radius from the barycentre of the binary, $a_{\mathrm p}$ is the planet semimajor axis and the power-law index $\gamma=1$.  Initial positions in the vertical ($z$) direction were similarly generated by randomly sampling the density distribution
\begin{equation}
\rho(z) = \rho_0\,\mathrm{exp}\left(-\frac{z^2}{2H^2}\right)
\label{eq_rho}
\end{equation}
where $\rho_0$ is the midplane density and $H$ is the scale-height of the disc.
We use a locally isothermal equation of state where temperature $T$ is a function only of $R$, such that $T \propto R^{-1/2}$. This gives a flared disc with thickness $H/R \propto R^{5/4}$. Particles were initially given zero velocity in the $z$ direction, and azimuthal values $v_{\phi}$ such that
\begin{equation}
\frac{v_{\phi}^2}{R} = \frac{G M_{\mathrm{b}}}{R^2}+\frac{1}{\rho}\frac{\upartial P}{\upartial R},
\label{eq_v}
\end{equation}
where $M_{\mathrm{b}}$ is the total binary mass and $P$ is the local pressure in the disc.

In order to reduce the impact of numerical transients, we allowed our discs to relax for a period of 100 binary orbits before inserting the planet into the simulation. Although the initial surface density profile has $\gamma = 1$, by the time the disc had relaxed this had become $\gamma \sim 3/2$, as expected for a circumbinary decretion disc \citep{pringle91}. In addition, in order to prevent the planet undergoing a transient burst of accretion immediately following its insertion, we modified the (relaxed) surface density profile to include a gap at the initial orbital semimajor-axis of the planet, using the parameterization of \citet{lubowdangelo06}, before inserting the planet.

\subsection{Model parameters}\label{model parameters}
\begin{table}
\begin{minipage}[t]{\columnwidth}\centering
\caption{Table of model parameters. $M_p$ is the planet mass in units of the mass of Jupiter, \Mjup, and $e_0$ is the initial eccentricity of the planet's orbit.}
\label{table1}
\begin{tabular}{lcl}
\hline
Model name & $M_p$ [\Mjup] & $e_0$ \\
\hline
{\sc reference} & 0.333 & 0.0 \\
{\sc massive} & 1.0 & 0.0 \\
{\sc eccentric} & 1.0 & 0.05 \\
\hline
\end{tabular}
\end{minipage}
\end{table}

We ran a total of 3 such simulations\footnote{Animations of these simulations can be seen and downloaded from \url{http://www.astro.le.ac.uk/users/acd23/K16.html}} for a period of 400 orbits of the planet (approximately 2200 binary orbits). In each run the physical parameters of the binary are those reported by \citet{doyleetal11}, and both the planet and disc were coplanar with the binary. The sink radii of the binary components was set to 0.25 in code units (i.e., well inside the disc's inner cavity), and the sink radius of the planet was set to be 0.4 of its Hill radius $R_{\mathrm{Hill}} = a_{\mathrm{p}} (M_p/3M_{\mathrm{b}})^{1/3}$. The initial semimajor axis of the planet was $a_{\mathrm{p}} = 0.7048$ au ($3.142$ in our code units) as reported in the discovery paper, and the initial radial extent of the disc (before relaxation) was $1.5 < R < 10$ in code units. We normalised $H$ so that it was 0.05 at a radius $R$ equal to the planet's semimajor-axis $a_{\mathrm{p}}$. With $10^7$ SPH particles we resolve the disc scale height into approximately 6 SPH smoothing lengths at the radius of the planet's orbit, indicating that we are resolving the vertical structure adequately to avoid numerical problems. The surface density was normalized such that $\Sigma_{\mathrm{p}} = 100$ g\,cm$^{-2}$ (though the actual surface density at $R=a_{\mathrm p}$ is much smaller than this, due to the gap in the disc at the planet's orbit).

Our {\sc reference} model used a planet mass $M_p= 0.333$\Mjup, as reported by \citet{doyleetal11}. To test the effect of a higher planetary eccentricity, we also ran a model with initial eccentricity $e_0 = 0.05$ ({\sc eccentric}). In this case we used a higher planet mass of 1 \Mjup\, to ensure that we did not see unphysical numerical eccentricity damping \citep[see e.g.][]{massetogilvie04}. In order to compare this run with our reference model, we also ran a model identical to the reference model but with this higher planet mass ({\sc massive}). The planet (along with the binary components) was modelled as an N-body particle and thus its orbital elements were free to evolve throughout the simulation. Other model parameters which were varied between our simulations are detailed in Table \ref{table1}. Our simulations were run on the DiRAC2 {\it Complexity} cluster\footnote{See \url{http://www.dirac.ac.uk/}} on up to 128 parallel cores, using approximately 300,000 core hours per run. 

\section{Results}\label{results}

\begin{figure}
\includegraphics[width=\columnwidth]{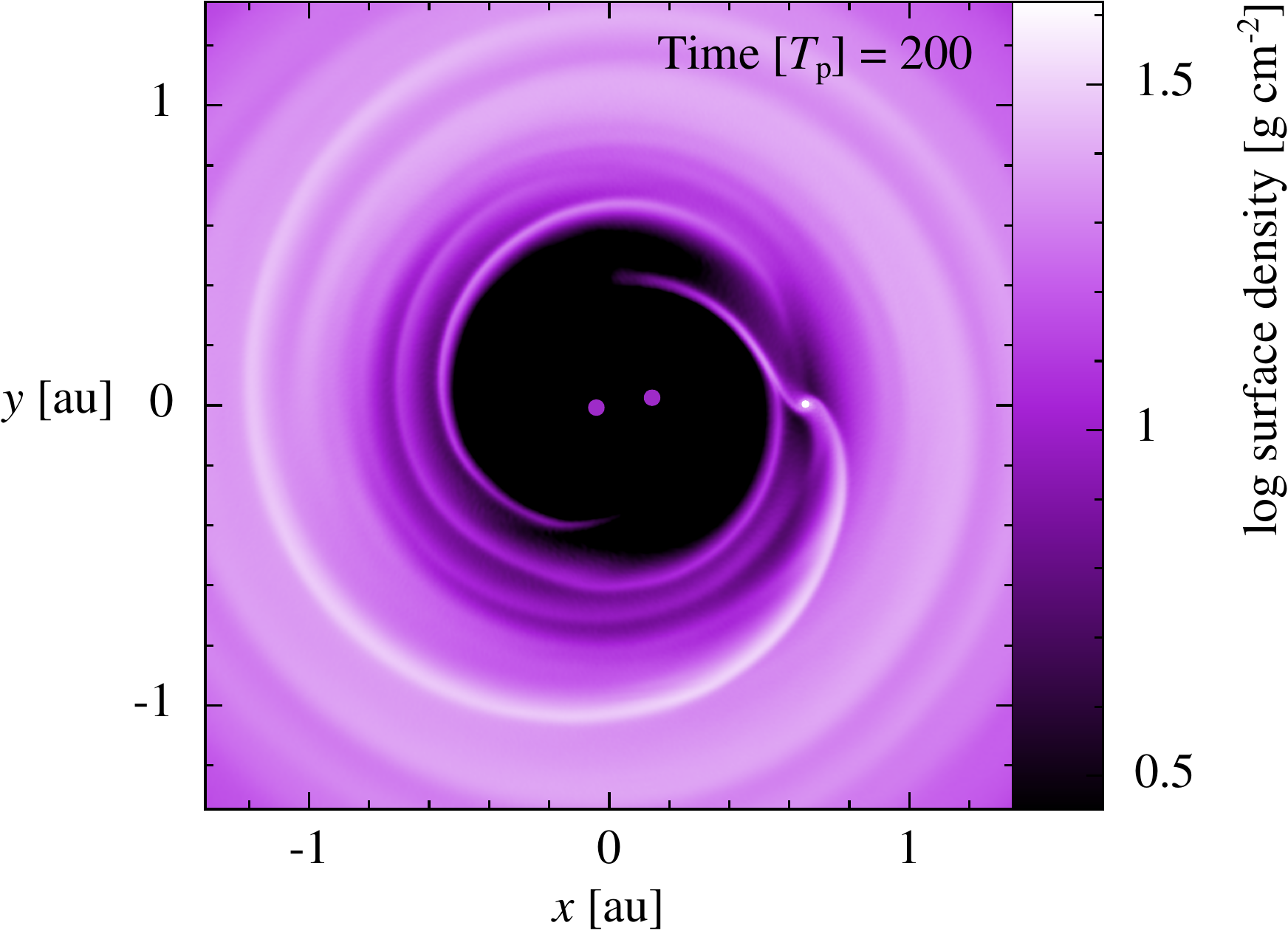}
\caption{Snapshot of the central region of our \textsc{massive} model (see Table \ref{table1}) after 200 planetary orbits. The positions of the (stellar) binary components are shown as purple points, while the white point denotes the position of the planet; the colour-scale shows the gas surface density. The planet orbits very close to the inner disc edge, and the spiral density waves induced by the planet are clearly visible. (Figure rendered using \textsc{splash}; \citealt{price07}.)}
\label{fig_disc}
\end{figure}

A representative snapshot of the central region of one of our simulations is shown in Figure \ref{fig_disc}. The trend in all of our models (shown in Figure \ref{fig_eccs}) follows the pattern found by \citet{leunglee13}, where the osculating eccentricity $e_{\mathrm{p}}$ consists of forced and free components. They found that this occurs between $e_{\mathrm{min}} = | e_{\mathrm{forced}} - e_{\mathrm{free}}| \simeq 0.006$, and $e_{\mathrm{max}} = e_{\mathrm{forced}} + e_{\mathrm{free}} \simeq 0.066$.  We see good agreement with this in our models with no initial eccentricity, for both low and high planet masses, suggesting that torques from the disc do not alter the planet's orbit significantly on dynamical time-scales.  In our initially eccentric model we see a similar osculating eccentricity, but with a larger amplitude (as expected), giving $e_{\mathrm{max}} \simeq 0.12$ for $e_0 = 0.05$.

As the planet's eccentricity evolution is  dominated by the binary forcing pattern, it is difficult to determine if there is any underlying longer-timescale change in the planet's eccentricity due to the interaction with the disc. In order to test this, we have compared the eccentricity in our simulations with that from a pure $N$-body run (that is, the same set-up as in our \textsc{reference} model but with no circumbinary disc). This is shown in Figure \ref{fig_EoverE}, where we plot the orbit-averaged \textsc{reference} planet eccentricity ($e_{\mathrm{hydro}}$) divided by the orbit-averaged $N$-body planet eccentricity ($e_{N\mathrm{-body}}$).  Relative to the N-body model \citep[which closely matches the analytic solution of ][]{leunglee13}, the \textsc{reference} model shows initial eccentricity damping (at the per cent level) due to the relaxation of the disc initial conditions, followed by periodic oscillations.  This periodicity matches the osculation period seen in Figure \ref{fig_eccs}, and consequently we do not attach great significance to the minima seen in Figure \ref{fig_EoverE} (as these represent ratios of pairs of very small numbers).  The maxima, however, are more reliable, and after the decay of initial transients these eccentricity peaks show a shallow decline with time. Due to the orbital precession induced by the disc, the eccentricity cycles from the \textsc{reference} model and the $N$-body run are slightly out of phase by the end of our simulations.  This causes the `sharpening' of the maxima seen in Figure \ref{fig_EoverE}, but the phase difference is sufficiently small that the peak values are not strongly affected. We therefore conclude that the disc damps the planet eccentricity (relative to the N-body run) at the 1 per cent level over the duration of our simulations.

\subsection{Torque analysis}\label{torques}
\begin{figure}
\includegraphics[width=\columnwidth]{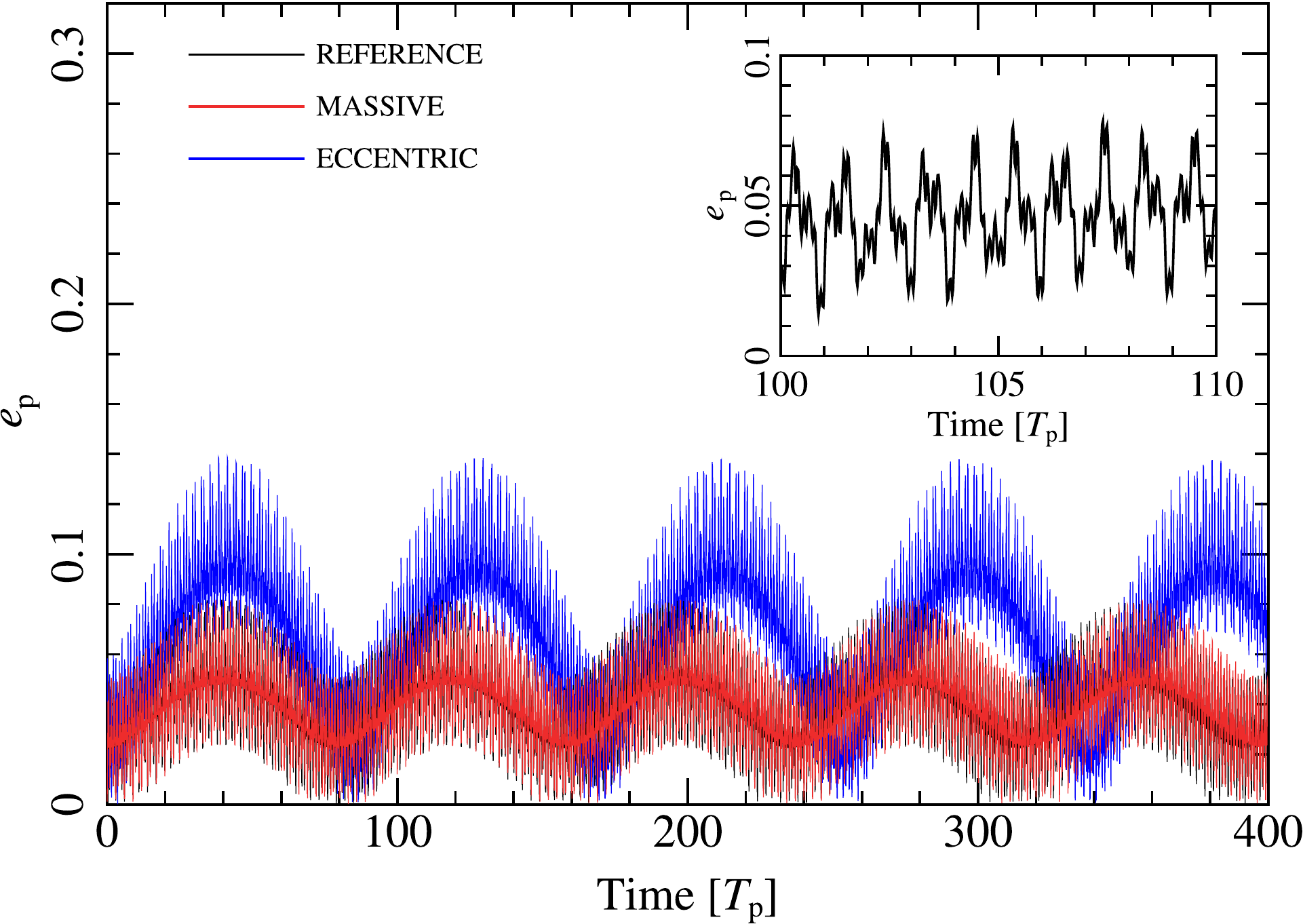}
\caption{Planet eccentricity as a function of time for the three models described in Table \ref{table1}. The osculation of eccentricity for our initially circular models ({\sc reference} and {\sc massive}) agrees with those found by \citet{doyleetal11} and with the analytic theory of \citet{leunglee13}, independent of planet mass. The inset shows the eccentricity over 10 orbits in the {\sc reference} model, illustrating the forcing pattern driven by the central binary.}
\label{fig_eccs}
\end{figure}

\begin{figure}
\includegraphics[width=\columnwidth]{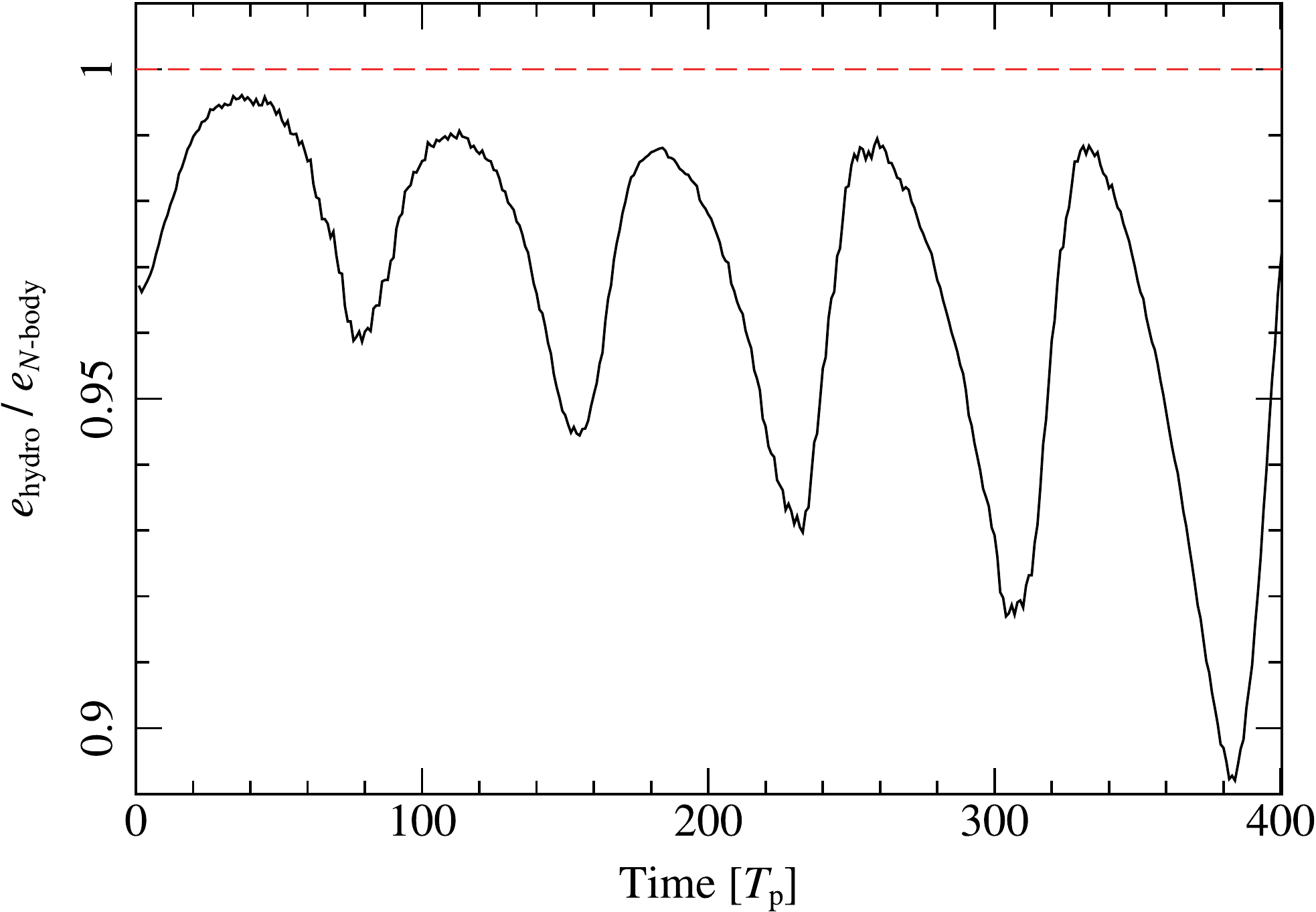}
\caption{Comparison between the orbit-averaged planet eccentricity from our \textsc{reference} run ($e_{\mathrm{hydro}}$) with that from an identical simulation with no gas disc ($e_{N\mathrm{-body}}$). For reference, the red dashed line shows denotes $e_{\mathrm{hydro}} / e_{N\mathrm{-body}} = 1$. We see strong oscillations due to the binary forcing, but after initial transients have dissipated the peak eccentricity settles into a broadly constant rate of decay.}
\label{fig_EoverE}
\end{figure}

The planet's orbit is dominated by the binary forcing throughout our simulations, as discussed above, but we can gain additional insight into its behaviour on longer time-scales by looking at the torques exerted on the planet by the disc.  Figure \ref{fig_TQs} shows the orbit-averaged disc torques on the planet for each of our simulations. The torque $\Gamma$ is defined such that a negative torque corresponds to eccentricity growth \citep{armitage10}; positive torques damp the planet's eccentricity.  Although the details vary between models, the same broad trend is seen in all our simulations.  The torques undergo periodic oscillations on the time-scale of the eccentricity osculation seen in Figure \ref{fig_eccs}, with the maximum torque corresponding to the peak eccentricity.  The underlying trend, however, is the torques are initially negative, but increase with time and eventually settle into a quasi-steady (oscillating) state. In the case of our \textsc{reference} model, this has a net positive torque, with a time-averaged value of $\langle\Gamma_{\mathrm {disc}}\rangle \simeq 6 \times 10^{36}$g\,cm$^2$s$^{-2}$. In a sense this is to be expected: the planet orbits near the inner edge of the disc, and the presence of more material exterior to the planet's orbit favours positive torques. These provide angular momentum to the planet, in principle allowing it either to reduce its eccentricity or to migrate outwards. However, our model setup assumes that the planet has previously migrated from a formation point further out in the disc, and this configuration disfavours outward migration, so the long-term effect of these positive torques is to damp the planet's eccentricity.

Our \textsc{massive} and \textsc{eccentric} models show the same basic trend of torque growth but here the net torque at the end of the simulation runs is negative. This is consistent with the results of \citet{pierensnelson08}, who found generally negative torques in their 2D disc models that resulted in the growth of eccentricity. Both our \textsc{massive} and \textsc{eccentric} models and the Saturn-mass run from \citet{pierensnelson08} were in the gap-opening regime due to high-mass planets and low disc viscosity respectively (see Figure \ref{fig_middens} for a comparison between our \textsc{reference} and \textsc{massive} models with respect to their gap opening). In contrast, our \textsc{reference} model is not fully in the gap-opening regime, and so feels additional torques from co-orbital gas not present in the \textsc{massive}, \textsc{eccentric} or \citeauthor{pierensnelson08}'s runs. We address numerical concerns with simulating non-gap opening planets in isothermal discs in Section \ref{limitations}.

In all of our simulations, the absolute values of $\Gamma$ are smaller than the tidal torques from the binary by a factor $\sim10^4$, and additional calculations (run for shorter durations) show that the disc torques scale linearly with $\Sigma$ over a wide range (at least 3 orders of magnitude) in disc surface density. We therefore conclude that if the planet is not in the gap-opening regime, the disc-planet interaction in the Kepler-16 system generally leads to damping of the planet's free eccentricity, with the damping time-scale determined primarily by the local disc surface density.

\subsection{Disc properties}\label{properties}

\begin{figure}
\includegraphics[width=\columnwidth]{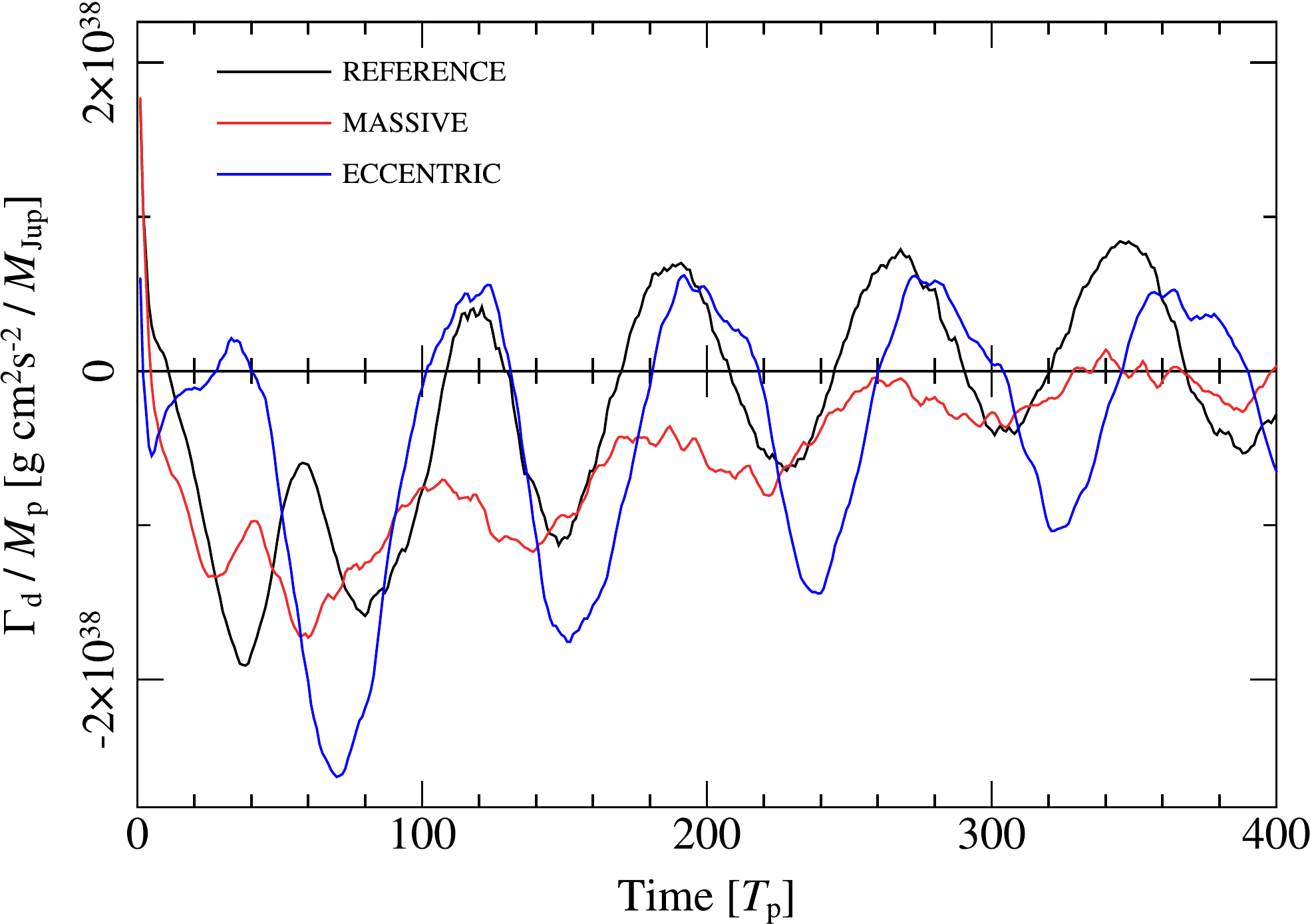}
\caption{Orbit-averaged disc torques, normalised to the planet mass, as a function of time (in planetary orbital periods) for the models described in Table \ref{table1}. The torques oscillate on the same time-scale as the eccentricity, and after initial transients have decayed settle into a pattern with a net positive torque of $6 \times 10^{36}$ g cm$^2$s$^{-2}$ (for our {\sc reference} model).  The torques on the planet from the binary are $\sim10^4$ times stronger, but additional runs show that the disc torques scale linearly with the surface density over a wide dynamic range in $\Sigma$. Lower-resolution test runs indicate that the torques in the \textsc{reference} model are approximately converged, and the \textsc{massive} and \textsc{eccentric} models are almost converged by the end of the simulation.}
\label{fig_TQs}
\end{figure}

We can now use this result to infer a limit on the gas surface density of the real disc in which Kepler-16b formed. Previous simulations of circumbinary planets have have found that eccentricity grows on a time-scale comparable to (or shorter than) the migration time-scale $\tau_{\mathrm{mig}}$ \citep[e.g.,][]{pierensnelson08} . If the planet's orbit never became significantly eccentric during its migration, then we require that the eccentricity be damped on a time-scale $\tau_{e} \lesssim \tau_{\mathrm{mig}}$.  In order to connect this to the disc torque, we must calculate the angular momentum $\Delta J$ gained by the planet from the damping torque. To first-order the potential is Keplerian, so we approximate the planet's angular momentum as $J_{\mathrm{p}} = M_{\mathrm{p}}\sqrt{GM_{\mathrm{b}}\,a_{\mathrm{p}}(1-e^2)}$.  We then differentiate to find
\begin{equation}
\left(\frac{\upartial J_{\mathrm{p}}}{\upartial e}\right)_{a_{\mathrm{p}}} = -M_{\mathrm{p}}\sqrt{GM_{\mathrm{b}}\,a_{\mathrm{p}}}\, \left(1-e^2\right)^{-1/2} e \,,
\end{equation}
so to first-order in $e$
\begin{equation}
\frac{\mathrm dJ}{\mathrm de} \simeq -e M_{\mathrm{p}}\sqrt{GM_{\mathrm{b}}\,a_{\mathrm{p}}}
\label{eq_djde}
\end{equation}
and
\begin{equation}
\Delta J = \left|\frac{\mathrm dJ}{\mathrm de}\right|e = e^2 M_{\mathrm{p}}\sqrt{GM_{\mathrm{b}}\,a_{\mathrm{p}}} \, .
\label{eq_deltaj}
\end{equation}
In order to maintain a low eccentricity we require a damping torque
\begin{equation}
\Gamma_{\mathrm{d}} \gtrsim \frac{\Delta J}{\tau_e} \, .
\label{eq_gamma}
\end{equation}
We find that $\Gamma_{\mathrm{d}}$ scales linearly with the disc surface density (in the limit $M_{\mathrm{d}} \ll M_{\mathrm{b}}$), as discussed above, so this allows us to place a lower limit on $\Sigma$.  The planet is unlikely to have attained an eccentricity higher than $e=0.1$ \citep[e.g.][]{popovashevchenko12}, which gives $\Delta J = 2.3 \times 10^{47}$ g cm$^2$s$^{-1}$.  For our {\sc reference} model, which has $\Sigma_{\mathrm{p}} = 100$ g\,cm$^{-2}$, the net damping torque is $\Gamma_{\mathrm{d}} \simeq 6 \times 10^{36}$ g\,cm$^2$s$^{-2}$ (see Figure \ref{fig_TQs}), so we can substitute these values into Equation \ref{eq_gamma} and re-arrange (assuming $\Gamma_{\mathrm{d}} \propto \Sigma_{\mathrm p}$) to find
\begin{equation}
\Sigma_{\mathrm p} \gtrsim 120 \, \left(\frac{\tau_e}{10^4\,\Omega_{\mathrm p}^{-1}}\right)^{-1} \mathrm g\,\mathrm {cm}^{-2} \, .
\end{equation}
This is a surprisingly large surface density, and implies that Kepler-16b formed in a massive circumbinary disc. \citet{pierensnelson08} found that the migration and eccentricity growth time-scale for a Saturn-mass planet is $\sim 10^4\,\Omega_{\mathrm p}^{-1}$, which we adopt as a reference value above.  The true migration time-scale is not known, and the planet may well have migrated more slowly.  However, even a conservative assumption of $\tau_e \sim 10^5\,\Omega_{\mathrm p}^{-1}$ requires $\Sigma_{\mathrm p} \gtrsim 10$ g\,cm$^{-2}$, and we regard this as a reasonably robust lower limit to the surface density of Kepler-16b's parent disc.

\section{Discussion}\label{discussion}

\subsection{Limitations of the model}\label{limitations}

A major source of uncertainty in our model is our treatment of the disc thermodynamics. We adopt a locally (vertically) isothermal equation of state, but in the single-star case it is well known that a full radiative hydrodynamical treatment can give different results for low-mass planets below the gap-opening limit \citep[e.g.][]{bitschkley10,bitschkley11a,bitschkley11b}. For Saturn-mass planets, it seems that migration behaviour is broadly unaffected by the disc equation of state \citep{bitschkley11a}, but the disc structure can still be strongly affected -- \citet{marzarietal13} compared their fully radiative treatment with the locally isothermal approximation of \citet{pelupessyportegieszwart13} and found significant differences, though the different initial surface density profiles also play a strong role.

\begin{figure}
\includegraphics[width=\columnwidth]{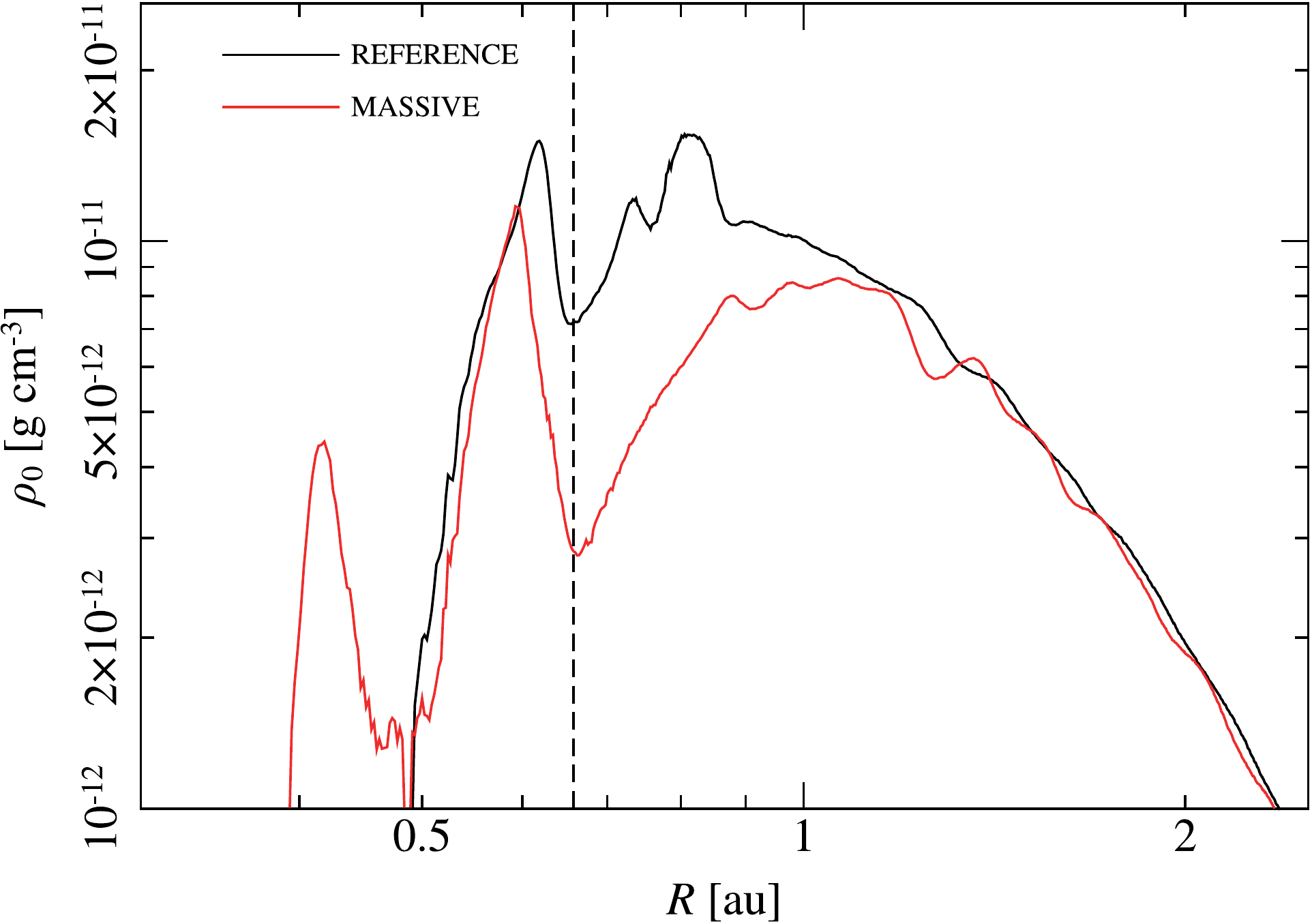}
\caption{Midplane gas densities for our {\sc reference} and {\sc massive} models (see Table \ref{table1}) after 200 planetary orbits, as a function of disc radius $R$ along an axis offset from that of the planet by 0.1 radians (to avoid the density enhancement of material being accreted by the planet). A dashed line indicates the semi-major axis of the planet in both models. While a Jupiter-mass planet is able to maintain a strong gap in the disc, the lower-mass planet in the {\sc reference} model only opens a partial gap, with a significant mass of gas remaining in the corotation region.}
\label{fig_middens}
\end{figure}

Unfortunately the parameter space for radiative circumbinary disc models in 3-D is vast, and exploring even a modest sub-set of this space is not feasible.  However, in single-star models the critical uncertainty where a gap is not opened is usually the (thermo)dynamics of gas in the co-rotation region (near the planet), but it is unclear if we are in this regime. We can investigate this in our simulations by comparing models with different planet masses. In Figure \ref{fig_middens} we plot the midplane density for the {\sc reference} and {\sc massive} models after 200 planetary orbits.  We see that the {\sc reference} model, with a planet of approximately Saturn mass, only opens a partial gap in the disc, and only the more massive planet is truly in the gap-opening regime. The results of our {\sc reference} model must therefore be taken with some caution. The torque analysis in Figure \ref{fig_TQs} shows that the torque oscillates on the same time-scale for all planet masses, but the net (time-averaged) torque is negative for the runs with higher planet mass, as seen in previous studies \citep[e.g.,][]{pierensnelson08}. However, \citet{bitschkley11b} showed in the single-star case that an accurate treatment of the disc thermodynamics generally increases the disc torques for low-mass non gap-opening planets, suggesting that the net positive $\Gamma_{\mathrm{d}}$ seen in our {\sc reference} simulation is a robust result. Clearly, further investigation of this issue is still required.

A further limitation of our high-resolution 3-D treatment is that due to computational expense, we are only able to simulate time-scales of hundreds of planetary orbits. In contrast, \citet{pierensnelson08} ran their 2-D simulations for tens of thousands of planetary orbits. Although we cannot follow changes in the planet's orbit on such long time-scales, our torque analysis provides a clean way to probe longer time-scale behaviour in our models.  However, it is notable that our result, that disc torques damp eccentricity, disagrees with \citet{pierensnelson08}, who found eccentricity growth and outward migration in comparable models.  It is well known that a 2-D approach can lead to artificially higher resonant torques \citep*{tanakaetal02}, but this is unlikely to be the origin of such a qualitative change in behaviour.

Instead, we attribute this apparent discrepancy to the different disc viscosities used.  We chose $\alpha = 10^{-2}$, while \citet{pierensnelson08} used $\alpha = 10^{-3}$.  The different dimensionality (3-D versus 2-D) means that we cannot compare the gap-opening criteria in these simulations directly, and it is computationally unfeasible for us to simulate such low disc viscosities \citep[see discussion in][]{dunhilletal13}.  However, the critical planet mass for gap-opening typically decreases with lower viscosities, as the disc-planet torques more easily overcome the viscous torques in the disc.  Comparing our Figure \ref{fig_disc} with Figures 5 and 8 in \citet{pierensnelson08}, we see that their planets open much deeper and wider gaps in their discs, and therefore qualitatively different behaviour is not unexpected. Measurements of real disc viscosities are uncertain at best, and are almost certainly not constant (particularly in the case of ``dead zones'' in disc midplanes).  In general, however, our adopted value of $\alpha = 10^{-2}$ is consistent with the values inferred from observations of protoplanetary discs \citep*[e.g.][]{hartmannetal98,kingetal07}. We note also that our simulations are specifically tailored to the observed parameters of the Kepler-16 system, and therefore differ from those of \citet[][who used $q=0.1$, and initial eccentricities of $e_{\mathrm{b}} = 0.08$ and $e_{\mathrm{p}} = 0.02$]{pierensnelson08}. Their more extreme binary mass-ratio, in particular, may change the system's dynamics significantly (as the secondary orbits much closer to the inner disc edge); further exploration of the binary parameter space is necessary to understand these effects in more detail.

The relatively short duration of our simulations means that we cannot rule out the possibility that the positive torques on the planet drive some outward migration in addition to damping the planet's eccentricity. However, as noted in Section \ref{torques}, we find this unlikely, as our simulations consider a planet that has already migrated to the orbit we see it at today. Figure \ref{fig_EoverE} shows that we do indeed see some small level of eccentricity damping in our simulations, although again we cannot rule out the possibility of outward migration at later times. If we do consider this as a possible outcome, it in fact strengthens our minimum surface density requirement. This is because if some fraction of the total disc torque is `diverted' towards driving outward migration, the remaining torque which acts to damp eccentricity is smaller, requiring a higher surface density to maintain the same level of eccentricity damping.

Our analysis also neglects the possibility that the binary can drive eccentricity in the disc, which in turn can alter the angular momentum of the planet. No significant disc eccentricity is seen in any of our simulations, but \citet{pierensnelson08} showed that a binary with mass ratio and eccentricity similar to Kepler-16 can excite a disc eccentricity of $e_{\mathrm{d}} \sim 0.05$ on time-scales only slightly longer than the migration time-scale of the planet (i.e., several thousand planetary orbits). If we consider the limiting case where the planet absorbs all of the additional angular momentum within $\pm H$ of its orbit, we find that a disc eccentricity of 0.05 corresponds to $\simeq 10$\% of the planet's angular momentum for $\Sigma_{\mathrm{p}} \simeq 400$g cm$^{-2}$. Detailed consideration of this effect is beyond the scope of our simulations, but this suggests that disc eccentricity is unlikely to alter our conclusions significantly unless the disc is very massive. Instead, as the time-scale for disc eccentricity growth is relatively long, the most likely outcome is that a modest disc eccentricity will slow, but not reverse, the damping of the planet's eccentricity. 

Finally, we note that our simulations and analysis are predicated on the assumption that Kepler-16b's eccentricity was damped during the disc phase, rather than at later times in the systems evolution.  This is supported by dynamical modelling of the system, which suggests that the planet's eccentricity never exceeded $0.05$ in the post-disc phase \citep{popovashevchenko12}, but it is possible that an additional body (or bodies) in the system could influence the planet's orbit.  The simplest dynamical explanation for the low eccentricity is that the planet is at the low-$e$ point of a Kozai cycle \citep{kozai62,lidov62}.  This scenario is rather contrived, however, and requires the presence of a massive perturber on an inclined orbit in addition to the system being observed at a very specific time in its evolution.  There is some evidence for an additional planetary-mass body in the system \citep{benderetal12}, but detailed dynamical modelling of its potential effects is not yet possible.  We therefore cannot exclude a dynamical origin, but given the level of fine-tuning required we do not consider this a likely explanation for the low eccentricity of Kepler-16b.

\subsection{Implications for circumbinary planet formation}\label{implications}
Our results suggest that Kepler-16b formed in, and migrated through, a circumbinary disc with a relatively high surface density.  Observations currently tell us little about the surface densities of circumbinary discs, particularly for young binaries with separations as close as Kepler-16.  At larger separations ($a \gtrsim 10$ au), where young binaries are more readily observed, most circumbinary discs appear to be rather low mass: the majority of the circumbinary discs identified by \citet{krausetal11,krausetal12} have masses $M_{\mathrm{d}} \leq 10^{-3}\,$ \Msun\,\citep[][]{andrewswilliams05}. At the other end of the binary period distribution, \citet{rosenfeldetal12} measured the mass of CO gas in the disc surrounding the very close (2.4-day period) binary V4046 Sgr to be $M_{\mathrm{CO}} \sim 3 \times 10^{-6}\,$\Msun, implying a total gas mass of $\sim 0.01-0.1$ \Msun. This system is particularly noteworthy because it is old \citep[$\sim 12$ Myr;][]{rosenfeldetal12}, yet it still retains a very massive disc.

However, all of these measurements are based on (sub-) millimetre observations, which only probe the outer regions of the disc, and estimates of disc surface densities at au radii can only be made by extrapolating inwards. Identifying young binaries at au separations remains challenging \citep[e.g.,][]{krausetal12} and, while some subset of our sample of circumstellar discs are presumably circumbinary, observations of discs in close binary systems remain scarce. Our limit on the surface density of Kepler-16b's parent disc therefore provides a useful new insight into circumbinary disc physics in this poorly-explored regime.

\begin{figure}
\includegraphics[width=\columnwidth]{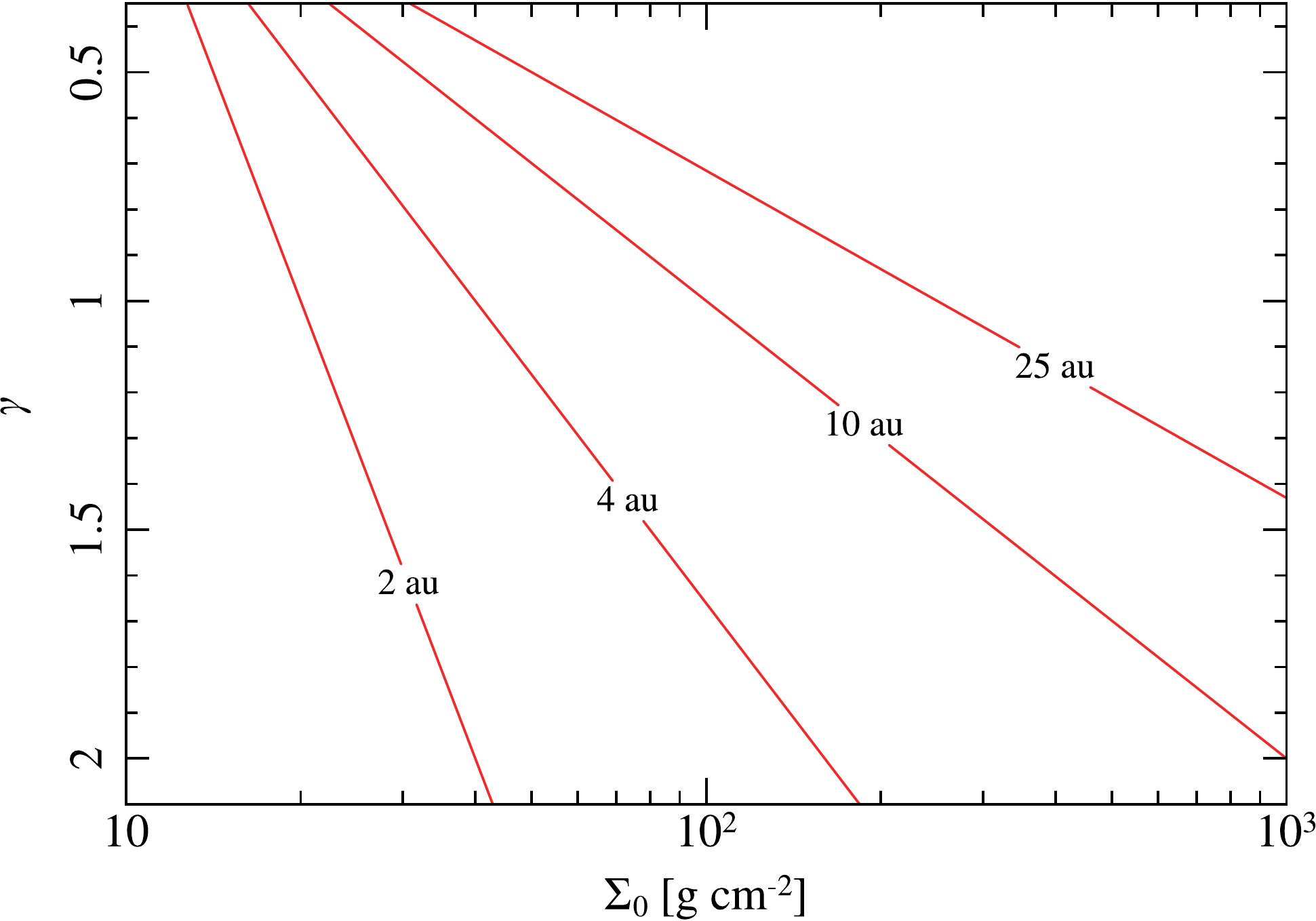}
\caption{Contour plot for disc models the form of equation \ref{eq_sigma2}. Contours are labeled with the radius $R$ at which the surface density $\Sigma(R) = \Sigma_{\mathrm{min}}$ for given values of the normalisation factor $\Sigma_0$ and power-law index $\gamma$.}
\label{fig_contours}
\end{figure}

We extend our analysis further by considering the radial distribution of mass in such a disc. Figure \ref{fig_contours} shows a contour plot for different disc models with the form
\begin{equation}
\Sigma(R) = \Sigma_0\,\left(\frac{R}{1\,\mathrm{au}}\right)^{-\gamma},
\label{eq_sigma2}
\end{equation}
where contours indicate the radius $R$ at which the surface density equals $\Sigma_{\mathrm{min}} =$ \sigmin. The true radius at which Kepler-16b formed is unknown, although there is general agreement that it must have formed beyond approximately 10 au \citep{meschiari12,rafikov12,marzarietal13}. We can see from Figure \ref{fig_contours} that to satisfy $\Sigma(R \gtrsim 10\,\mathrm{au}) > \Sigma_{\mathrm{min}}$ we require discs with $\Sigma_0 \gtrsim 10^2$ g cm$^{-2}$, even for shallow surface density gradients ($\gamma < 1$).

We have focused our attention on Kepler-16b because of its extremely low measured eccentricity, but it in fact shares a number of characteristics with other circumbinary systems. Kepler-35b, -38b and -47b have low eccentricities ($e_{\mathrm{p}} < 0.05$) and similar semimajor-axis ratios $a_{\mathrm{p}}/a_{\mathrm{b}} \sim 3$--$4$. Moreover, the secular disc evolution models of \citet{alexander12} find that photoevaporation rapidly erodes the inner regions of circumbinary discs, and high surface densities persist only for a short fraction of the discs' lifetimes ($\lesssim 1$Myr).  Taken together with our results, this suggests that these circumbinary planets must have formed early in the evolution of their parent discs, when the discs were still massive enough to damp their eccentricities.  At present the poor observational statistics for both circumbinary discs and circumbinary planets do not allow us to draw detailed conclusions regarding the frequency of such planets, but as our census of both discs and planets improves we expect circumbinary systems to continue to provide crucial insights into the physics of planet formation.

\section{Conclusions}\label{conclusions}
We have performed high resolution 3D SPH simulations of a Kepler-16-like system embedded in a circumbinary disc. Analysing the disc torques from these simulations and using observational constraints on the system's history, we conclude that the planet must have maintained its low eccentricity by migrating through this disc without opening a significant gap, leading to the damping of eccentricity by co-orbital material. As this damping is directly proportional to the disc mass, we can set a limit on the local disc surface density in which Kepler-16b was once embedded, $\Sigma_{\mathrm{min}} \sim$ \sigmin. Applying this result to similar circumbinary systems also discovered by the {\it Kepler} mission, we conclude that this low limit may be used to provide constraints on the route planet formation takes in the circumbinary environment. We argue that the process requires relatively massive circumbinary discs, and that their secular evolution of implies that these planets form far out in the disc at early times, and migrated inwards before the disc mass was significantly depleted.

\section*{Acknowledgments}
We thank Dimitri Veras, Steinn Sigur$\eth$sson, Sijme-Jan Paardekooper, Graham Wynn and Richard Nelson for useful discussions and comments. We also thank an anonymous referee for a number of useful comments. ACD is supported by an Science \& Technology Facilities Council (STFC) PhD studentship.  RDA acknowledges support from STFC through an Advanced Fellowship (ST/G00711X/1). 
Theoretical Astrophysics in Leicester is supported by an STFC Consolidated Grant (ST/K001000/1).  
This research used the ALICE High Performance Computing Facility at the University of Leicester. Some resources on ALICE form part of the DiRAC Facility jointly funded by STFC and the Large Facilities Capital Fund of BIS. The project also made extensive use of the {\it Complexity} HPC cluster at Leicester which is part of the DiRAC2 national facility.

\bibliography{mnrasmnemonic,references}
\bibliographystyle{mnras}

\label{lastpage}

\end{document}